\documentstyle[psfig]{l-aa}                                    

\def\ros{{\sl ROSAT }}
\def\asca{{\sl ASCA }}

\def\G{$\Gamma_{\rm x}$ } 
\def\NH{$N_{\rm H}$}

\def\degs{\ifmmode ^{\circ}\else$^{\circ}$\fi}
\def\arcmin{\ifmmode ^{\prime}\else$^{\prime}$\fi}
\def\arcsec{\ifmmode ^{\prime\prime}\else$^{\prime\prime}$\fi}
\def\approxlt{\mathrel{\hbox{\rlap{\lower.55ex \hbox {$\sim$}}
        \kern-.3em \raise.4ex \hbox{$<$}}}}
\def\approxgt{\mathrel{\hbox{\rlap{\lower.55ex \hbox {$\sim$}}
        \kern-.3em \raise.4ex \hbox{$>$}}}}
\begin{document}
 
   \thesaurus{03         
              (11.01.2;  
               11.09.1;  
               11.17.2;  
               11.19.1;  
               13.25.2)  
}

   \title{Properties of {\itshape{dusty}} warm absorbers and the case of \\ 
           {\large{\bf IRAS}}\,17020{\large{$+$}}4544} 
   \author{Stefanie Komossa \inst{1}, Norbert Bade \inst{2} } 

   \offprints{St. Komossa, \\ 
            skomossa@xray.mpe.mpg.de} 
 
  \institute{Max-Planck-Institut f\"ur extraterrestrische Physik,
             85740 Garching, Germany     \and 
             Hamburger Sternwarte, Gojenbergsweg 112, 21029 Hamburg, Germany 
       }

   \date{Received 4 November 1997; accepted 9 December 1997}

   \maketitle\markboth{St. Komossa, N. Bade ~~~ Properties of dusty warm absorbers}{}  

   \begin{abstract}
We present a study of the properties of {\em dusty} warm absorbers 
and point to
some important consequences for the resulting X-ray absorption spectra
of AGN.    
Pronounced effects of the presence of {\em dust} in {\em warm} material
are (i) an apparent `flattening' of the observed X-ray spectrum in the \ros band
due to a sequence of absorption edges and a shift to lower gas ionization,  
(ii) the presence of a strong carbon edge at 0.28 keV, 
and (iii) the expectation of increased time variability 
of the warm absorber parameters. 
The first two effects can be drastic and may completely hamper an X-ray spectral fit 
with a dusty warm absorber even if a dust-{\em free}
one was successfully applied to the data.  
In order to demonstrate facets of the dusty warm absorber model and
test the recently reported important, albeit indirect, evidence for dusty warm 
material in the Narrow-Line Seyfert 1 galaxy  
IRAS 17020+4544 we have analyzed \ros PSPC and HRI observations of this galaxy.  
The X-ray spectrum can be successfully described by a single powerlaw with index \G=--2.4
plus small excess cold absorption, or alternatively by a steeper intrinsic powerlaw
(\G $\simeq$ --2.8) absorbed by a {\em dusty} warm absorber.
The findings are discussed in light of the 
NLSy1 character of IRAS 17020 and consequences for NLSy1s in general 
are pointed out. In particular, 
the presence of dusty warm gas results in a steeper intrinsic powerlaw than observed, thus exaggerating
the `FeII problem'. It also implies
weaker potential warm-absorber contribution
to high-ionization Fe coronal lines.
 

   \end{abstract}
 
\section{Introduction}

Warm absorbers reveal their presence by imprinting absorption edges on the soft X-ray spectra
of active galaxies (AGN). 
Many were found and studied 
on the basis of observations by
\ros (e.g., Nandra \& Pounds 1992, Komossa 1997) and \asca (e.g., Mihara et al. 1994, Reynolds 1997). 
Within the last year evidence has accumulated that some warm absorbers
contain significant amounts of dust. This possibility was first suggested by Brandt et al. (1996) 
to explain the lack of excess X-ray {\em cold} absorption despite strong optical
reddening of the quasar IRAS 13349+2438.  
 
Models that explicitly include the dust-gas and dust-radiation 
interaction using the  photoionization code {\em Cloudy} (Ferland 1993) 
were calculated by Komossa \& Fink (KoFi hereafter; 1996, 1997a-d). 
Depending on the warm absorber parameters, the presence
of dust turned out to have a strong influence on the X-ray absorption structure.
The models were applied to several Seyfert galaxies. 
NGC 3227 (KoFi 1996, 1997b) and NGC 3786 
(KoFi 1997c) were shown to be very good candidates
for dusty warm absorbers as judged from optical-UV reddening properties as well as  
{\em successful X-ray spectral fits}. On the other hand, the bulk of the 
warm material in NGC 4051 was 
found to be dust-free (KoFi 1997a). 
Recently, Reynolds et al. (1997) also presented a {\em Cloudy}-based dusty
absorber model which they successfully applied 
to the outer warm absorber in MCG 6-30-15.
Indirect evidence for the association of some warm absorbers with dust was discussed
in Reynolds (1997). 

IRAS 17020, which is part of the present study, 
is a Narrow Line Seyfert\,1 galaxy (NLSy1; Moran et al. 1996, 
Wisotzki \& Bade 1997) at redshift $z$=0.06 (de Grijp et al. 1992). 
Wisotzki \& Bade stressed the heavy reddening of the optical spectrum.  
They also presented a powerlaw fit to the \ros PSPC spectrum
and pointed to the discrepancy between the cold column density derived from optical
reddening and the one from the X-ray fit.  
Leighly et al. (1997; L97) detected an oxygen OVII edge in the \asca
spectrum of IRAS 17020 as well as 
high optical polarization. They also confirmed the strong optical
reddening and derived a corresponding gaseous column density of 
$N_{\rm opt}$ = 4 $\times~10^{21}$ cm$^{-2}$. On this basis
they suggested the presence of a warm absorber with internal dust in IRAS 17020.  

To test whether the model of a warm absorber that {\em includes} the presence
of dust consistently fits the X-ray spectrum of IRAS 17020, we apply 
such a model to the (archival) \ros PSPC spectrum of this source (Sect. 2).  
Motivated by increasing indirect evidence for the 
presence of dusty warm absorbers, in the main part (Sect. 3) 
we point out some important general properties of the dusty material 
that can result in strong modifications of the X-ray absorption structure 
as compared to the dust-{\em free} case, and study consequences. 
Since the presence of dust influences the X-ray spectrum most
strongly at soft energies,  
one signature 
being a pronounced carbon edge at
0.28 keV outside the \asca sensitivity range, 
\ros (Tr\"umper 1983) is particularly well suited for such a study.  
To search for time variability of IRAS 17020, and check for the potential 
contribution to the X-ray spectral complexity
of this source from the closeby optically bright star SAO 46462, 
we also present (PI) HRI observations. 

\section{Data reduction, temporal and spectral analysis}

\begin{figure}[h]
      \vbox{\psfig{figure=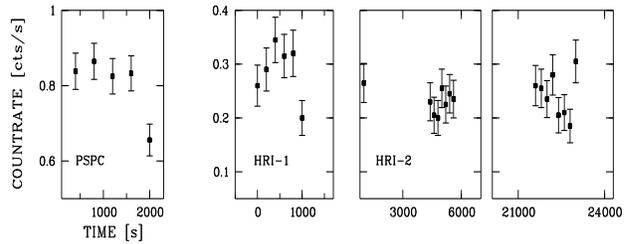,width=8.5cm,height=3.5cm,
          bbllx=2.8cm,bblly=7.1cm,bburx=18.0cm,bbury=12.2cm,clip=}}\par
      \vspace{-0.4cm}
 \caption[light]{PSPC and HRI X-ray lightcurve of IRAS 17020.
The time
is measured in seconds from the start of the individual observations. 
   }
 \label{light}
\end{figure}

\noindent {\em{PSPC}}.  
A 2.6 ksec pointed observation centered on IRAS 17020 was performed  
on Aug. 28, 1992. 
The background corrected source photons were extracted
and vignetting
and dead-time corrections were applied  
using EXSAS (Zimmermann et al. 1994).
The mean source count rate is 0.80$\pm{0.02}$ cts/s and there are hints
for 25\% variability during the observation (Fig. \ref{light}).

\vspace*{0.15cm} 
\noindent {\em{HRI}}. 
IRAS 17020 was observed with the HRI on March 4 (1.4 ksec), and April 4-5, 1994 (4.1 ksec).
We find mean source count rates of 0.29$\pm{0.02}$ and 0.22$\pm{0.01}$ cts/s, respectively. 
Again, there are hints for short-timescale variability (Fig. \ref{light};
we note that there are no strong variations in the background during the
observations).
The F8 star SAO 46462 in 108\arcsec~distance from IRAS 17020 is detected 
in X-rays. 
However, in the PSPC (HRI),  
the star is a factor of $>$50 (80) weaker  
than the target source
and thus does not significantly confuse the spectrum of IRAS 17020.  

\vspace*{0.15cm}
\noindent The conversion of all count rates to \ros PSPC rates
(assuming constant spectral shape) 
yields values of $CR$ = 0.91, 0.80, 0.71, 0.93 and 1.1 cts/s for the \ros survey,
PSPC, HRI-1, HRI-2 and \asca (L97) observation, respectively, revealing small-amplitude
long-term variability.  
 
\begin{figure}[t]      
      \vbox{\psfig{figure=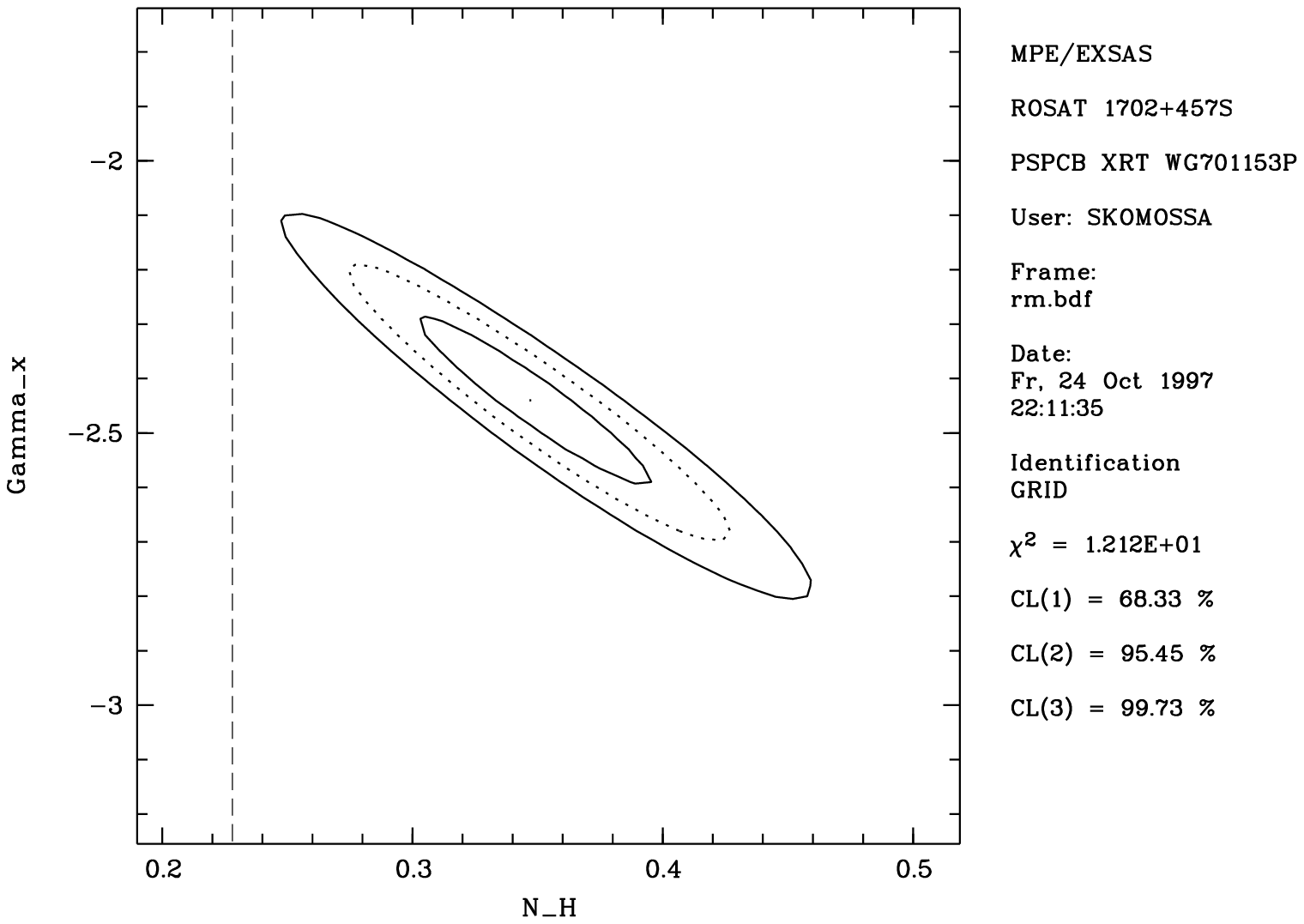,width=7.2cm,height=5.3cm,%
          bbllx=1.6cm,bblly=1.1cm,bburx=14.5cm,bbury=12.2cm,clip=}}\par
      \vspace{-0.4cm}
 \caption[chi]{Error ellipses  (confidence levels 
of 68.3, 95.5 and 99.7\%) in \G and \NH~(in units of 10$^{21}$ cm$^{-2}$)
 for the pl description of the
X-ray spectrum. 
The dashed line marks the Galactic absorbing column density
towards IRAS 17020.    }
 \label{chi}
\end{figure}

For the spectral analysis source photons in
the channels 11-240 were binned
according to a signal/noise ratio of 10$\sigma$.
A single powerlaw (pl) with cold absorption as a free parameter
gives a good fit to the PSPC X-ray spectrum ($\chi^2_{\rm red}$ = 0.8) 
with a photon index $\Gamma_{\rm{x}} = -2.4$, absorption 
of \NH~ = 0.35\,10$^{21}$ cm$^{-2}$ and a 1-keV photon flux $f$ = 2.71\,10$^{-3}$ ph/cm$^2$/s/keV.  
The error ellipses  
for this model are displayed in Fig. \ref{chi}.  

%
  \begin{figure}[thbp]
 \begin{center}
 \vbox{\psfig{figure=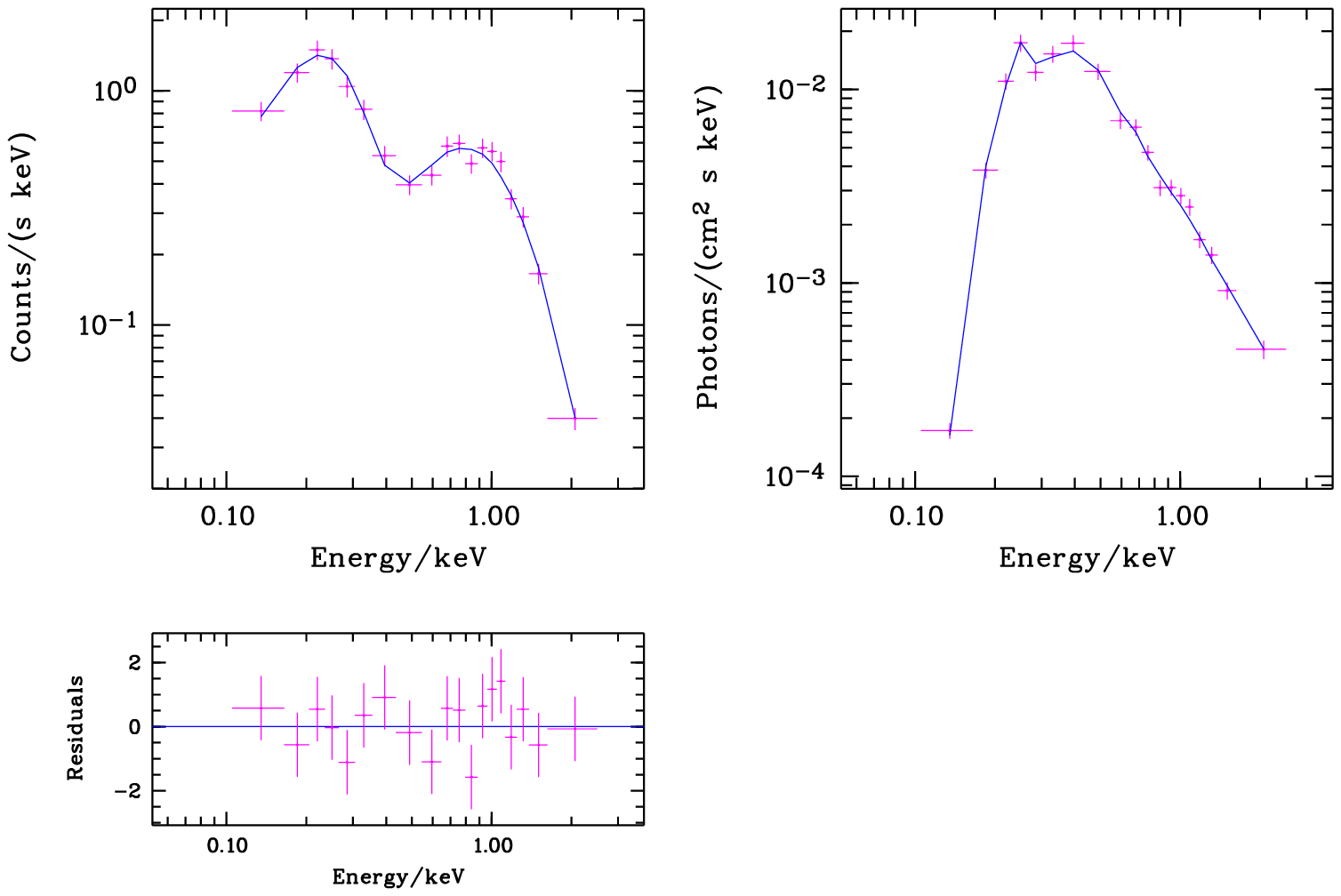,width=7.2cm,
          bbllx=1.9cm,bblly=1.2cm,bburx=10.1cm,bbury=4.4cm,clip=}}\par      
            \vspace{-0.45cm}
\caption[SEDx]{
Residuals for the fit of a dusty warm absorber model with \G=--2.8 and  
$N_{\rm w}$=$N_{\rm opt}$ ($\chi^2_{\rm red}$ = 0.77).  
  }
\label{SEDx}
\end{center}
\end{figure}

Application of other spectral models, like a black body or a Raymond-Smith model,
yielded no acceptable fits.
A pl plus black body and a double pl 
fit was performed, fixing the cold absorption to the value $N_{\rm opt}$ derived
from optical reddening. This was to test whether a low-energy soft excess
component could compensate for the stronger cold absorption thus explaining the data without
warm absorber.  
Again, no acceptable fit could be achieved.     
 
We have carefully searched for remaining systematic residuals in the pl fit
around the locations of the oxygen absorption edges -- none are found. 
To check whether a dusty warm absorber may nevertheless be present, with a sequence of
absorption edges not individually resolved (cf. Sect. 3), 
we have fit our {\em Cloudy}-based dusty warm absorber models
to the X-ray spectrum of IRAS 17020.  
The models and the chosen intrinsic continuum  
are described in more detail in Komossa \& Fink (1997a,b,c). In brief, the ionized material
was assumed to be of constant density, 
ionized by a `mean Seyfert' IR to gamma-ray continuum,
and in photoionization equilibrium (for cautious notes on the latter 
assumption see Krolik \& Kriss 1995).   
The dust composition and grain size distribution were chosen like 
in the Galactic diffuse interstellar medium (Mathis et al. 1977)
if not mentioned otherwise, as incorporated in
{\em Cloudy}, and the metal abundances
were depleted correspondingly
(see Ferland 1993, Baldwin et al. 1991 for details).   
The fit parameters of the dusty warm absorber are its column density
$N_{\rm w}$ and the ionization parameter $U=Q/(4\pi{r}^{2}n_{\rm H}c)$.
 
In a first step, the photon index was fixed to the value derived from the fit of a single pl,
\G=--2.4. Fitting a warm absorber with free $N_{\rm H}$, $N_{\rm w}$ and $U$, we are led
back to the limit of no warm absorption. A fit with $N_{\rm w}$ fixed to 
$N_{\rm opt}$ yields cold absorption consistent with the Galactic value 
but $\chi^2_{\rm red}$ = 1.6. The fit becomes acceptable if non-`standard' dust (i.e., silicate
only) is chosen ($\chi^2_{\rm red}$ = 1). Allowing for steeper intrinsic powerlaw
spectra, we find satisfactory fits also for dust including the graphite species
(cf. Fig. \ref{SEDx}). 
Fixing $N_{\rm w}$ = $N_{\rm opt}$, the spectrum can be successfully described by
parameters (\G, $N_{\rm H}$, $\log U$) $\simeq$ (--2.6, 0.3\,10$^{21}$, 1.0) and 
(--3.0, 0.45\,10$^{21}$, 0.4) and values inbetween for $\chi^2_{\rm red} \leq$ 1.    
Finally, we checked whether the spectrum could be reconciled with the canonical 
photon index of \G = {\mbox{--1.9}} and a {\em dust-free} warm absorber.  
In this case we find 
$\log N_{\rm w} \simeq$ 23.2 and $\log U \simeq$ 0.9, but $\chi^2_{\rm red}$ = 1.2.
Although highly ionized high column density material may provide another explanation
for the observed optical polarization via electron scattering, this model
predicts a strong OVIII edge (not observed by {\sl ASCA}; L97) and
leaves systematic residuals in the \ros fit.     

\section{Discussion}

\subsection{Properties of dusty warm absorbers}
\begin{figure*}[t]
      \vbox{\psfig{figure=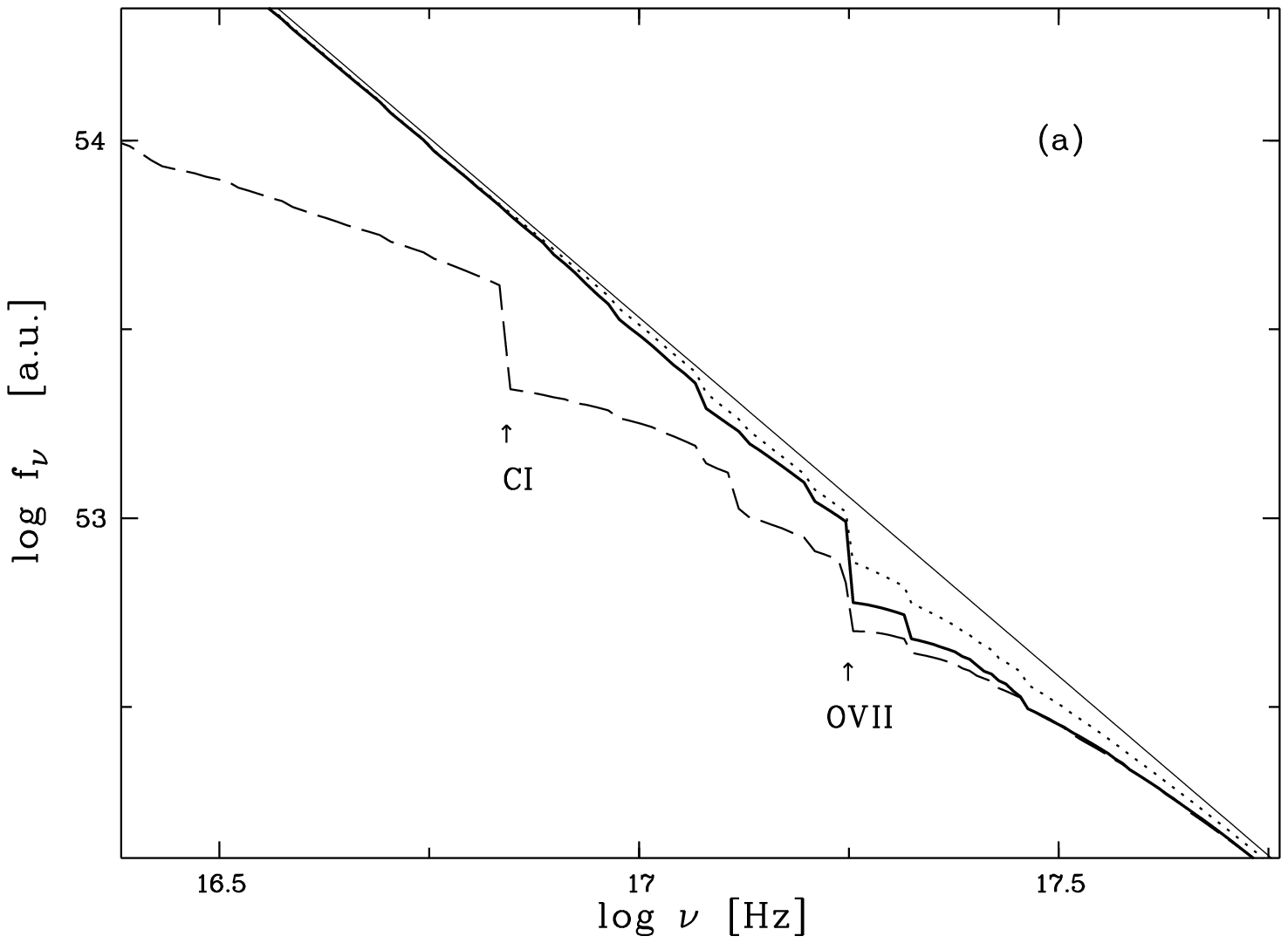,width=8.5cm,%
          bbllx=2.9cm,bblly=1.1cm,bburx=18.3cm,bbury=12.2cm,clip=}}\par
    \vspace*{-6.1cm}\hspace*{8.7cm}
      \vbox{\psfig{figure=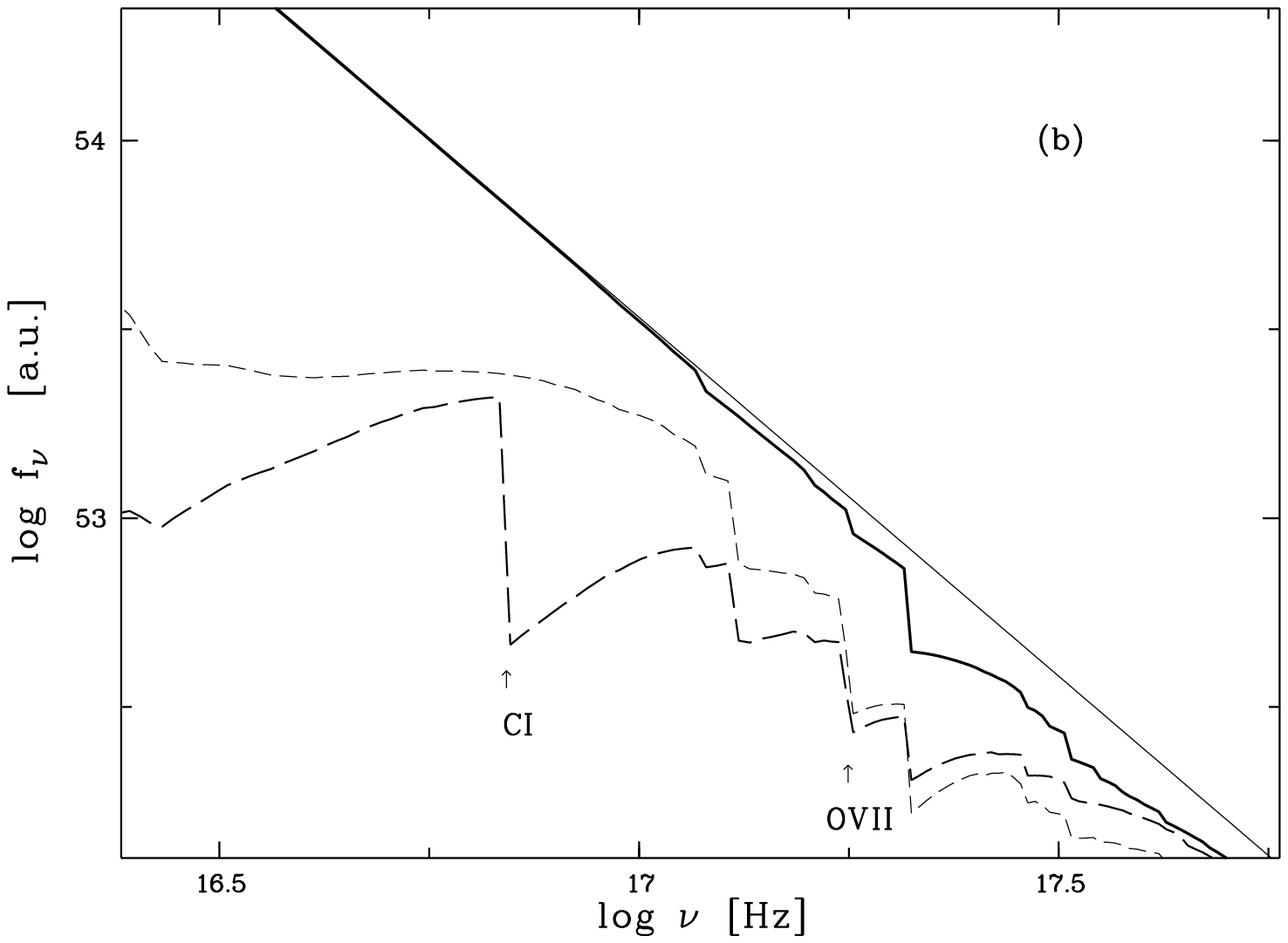,width=8.5cm,%
          bbllx=2.9cm,bblly=1.1cm,bburx=18.3cm,bbury=12.2cm,clip=}}\par
    \vspace*{-0.3cm}
 \caption[dust_seq]{Influence of dust mixed with the warm absorber on
       the X-ray absorption structure for two values of the warm column density
       ($\log N_{\rm w}$ = 21.6 and
       22.4).
       In {\bf (a)}     
       the straight solid line corresponds to the unabsorbed
       intrinsic spectrum, the heavy solid line to a dust-free warm absorber (WA) with solar abundances,
       the dotted line to a dust-free WA with the same depleted abundances as in the dusty model, 
       and the long-dashed line
       to the change in absorption structure after adding dust to the WA.
       In {\bf (b)}
       (in addition to the larger $N_{\rm w}$ a higher $U$ was chosen for presentation purposes)
       the heavy solid line again corresponds to a dust-free WA with solar abundances,
       the long-dashed line to a dusty WA (note that, for better presentation,
       the dust was even depleted by
       a factor $f_{\rm d}$ = 0.5), and the short-dashed line to the same dusty
       WA but without the graphite species of dust.
       Clear changes in the absorption structure are revealed for the models that include dust.
       The abscissa brackets the \ros energy range (0.1-2.4 keV); some prominent
       edges are marked. }
\label{dust_seq}
\end{figure*}

Some general properties of dusty warm absorbers are visualized in Fig. \ref{dust_seq}. 
The presence of dust modifies the X-ray absorption 
structure in two ways: \\
%
%
(i) Gas-dust interactions influence the thermal conditions in the gas and change
its ionization state (e.g. Draine \& Salpeter 1979). In particular,
for high ionization parameters dust effectively competes with the gas in absorbing
photons (Laor \& Draine 1993). 
Adding dust to large column density warm material (cf. Fig. \ref{dust_seq}b) results in strongly 
increased soft X-ray absorption, partially due to 
the stronger temperature gradient across the absorber, with more gas in a cold state. \\
(ii) Dust scatters and absorbs the incident radiation (e.g. Martin 1970)
and X-ray absorption edges are created by 
inner-shell photoionization of metals bound in the dust (cf. Figs 4,5 of Martin \& Rouleau 1991). 
Strong edges are those of neutral carbon and oxygen (Fig. \ref{dust_seq}a,b).
The shift in edge energies due to chemical and solid state effects is only of the order of a few eV (Greaves
et al. 1984).

The modifications of the observed X-ray spectrum as compared to the dust-free case 
can be drastic and are important to take into account when interpreting the observed X-ray properties 
of AGN. Consequences of the presence of dust are 
 
(A) `Flattening effect': The sequence of individual edges (dust-ones adding to the gas-ones) 
and the shift to lower ionization 
lead to a `smoothing'
and apparent `flattening' of the observed X-ray spectrum (Fig. \ref{dust_seq}a). 
Individual edges are less pronounced as in the dust-free case (where OVII and/or OVIII often
dominate). 
This `flattening effect' of dust (in fact opposite to that of a dust-free warm absorber,
which effectively steepens the X-ray spectrum in 
the \ros band) has consequences for NLSy1s in general (Sect. 3.3).  
Another consequence is the  
possibility to `hide' dusty warm absorbers in the \ros spectra (as shown by the successful fit to
IRAS 17020) if the column densities are not too large{\footnote {\small The
possibility of the presence of dusty warm material in a sample of NLSy1s 
to explain their polarization properties is 
suggested in Grupe et al. (1998).}},   
although effect (B) has to be taken care of (or else non-standard dust properties allowed;
dust in other galaxies may be different from Galactic one, but standard dust properties  
are also usually assumed to estimate optical reddening).

(B) Carbon edge:  Often, the strongest dust-created edge is that
of carbon at 0.28 keV, stemming from the graphite species of dust. 
Its presence may prevent a successful spectral fit and it is important to actually
apply the model of a dusty warm absorber to the data.  
Whereas dust including the graphite species  
is well consistent with the X-ray spectra of NGC 3227 (KoFi 1997b)
and NGC 3786 (KoFi 1997c) and pure graphite was favoured
by Reynolds et al. (1997) for MCG 6-30-15, 
it hampered a successful fit in other cases.

(C) Increased sensitivity to radiation pressure:
Material of high ionization parameter and particularly dust particles 
are strongly subject 
to radiation pressure (e.g., Laor \& Draine 1993, Chang et al. 1987, Binette et al. 1997), 
causing the gas to outflow (if not otherwise confined) and leading to temporal
changes of the absorber properties.  
Indeed, strong variability in X-ray count rate (of factors 10-15)
has been recently detected for the dusty-warm-absorber candidates
NGC 3227 and NGC 3786 which could be explained by 
variability in column density or ionization state of the warm absorber (KoFi 1997b,c).   

\subsection{The warm absorber in IRAS 17020} 

Given the potentially strong modifications of the X-ray
spectrum in the presence of {\em dusty} warm material,
the successful X-ray spectral fit 
lends further support to the suggestion (L97) of the presence of a dusty warm 
absorber in IRAS 17020.  
Although there is evidence for small excess cold absorption (e.g. Fig. 1)
no description of the soft X-ray spectrum
that involves the {\em large cold} column inferred from
optical observations{\footnote{\small Just to recall the assumptions that go into
the approach of relating optical reddening by dust with X-ray absorption by cold
gas: (i) optical and X-ray continuum reach the observer along the same path, (ii)
the extinction is not strongly variable with time, and (iii) the dust/gas ratio is
approximately the Galactic value; in modelling the warm absorber we assume (iv) its
ionization state to be dominated by photoionization and {\em Cloudy} to be applicable,
and again (v) dust properties similar to Galactic dust.}}   
could be found.

The successful fit of a dusty warm absorber requires a steep underlying powerlaw of index \G 
$\approx -2.8$ (the data are consistent with a flatter slope, if the graphite component of
dust is excluded). 

\subsection{NLSy1 nature of IRAS 17020} 
NLSy1s like IRAS 17020 are generally characterized by narrow broad lines, strong FeII emission and 
steep X-ray spectra (see Brandt et al. 1997 for a recent discussion).  
Several models were suggested to explain their apparently steep soft X-ray spectra,
like a strong soft excess on top of a flat powerlaw (e.g. Puchnarewicz et al. 1992)
or a (dust-free) warm-absorbed flat powerlaw (e.g. KoFi 1997d).
Recent observations reveal increasing X-ray spectral complexity,  
with often both, a warm absorber and a soft excess present.  
The presence of {\em dusty} warm material adds further to this complexity.
In particular, the effective
flattening of the observed spectrum implies a steeper intrinsic one,
thus exaggerating the problem of producing 
the strong FeII emission (that in standard models 
requires hard X-ray photons) in NLSy1s. 

The possibility of a warm-absorber contribution to the optical high-ionization lines 
in Seyferts and NLSy1s 
was studied by KoFi (e.g. 1997d, Fig. 6). Here, we note that in case the 
absorber is  {\em dusty}, weaker contribution to line emission is found. 
In particular, due to the binding of Fe in dust, negligible contribution to
high-ionization Fe coronal lines is expected.

Although we find hints for X-ray variability of IRAS 17020 of the order of 25\% on 
timescales from several hundred seconds to years, this amplitude is remarkably small 
as compared to some other reported cases of Seyfert and NLSy1 variability.  

\subsection{Concluding remarks} 

We have shown how the presence of dust in warm absorbers can lead to strong 
modifications of the observed X-ray spectrum. 
The model of a dusty warm absorber was successfully fit to the   
X-ray spectrum of the NLSy1 galaxy IRAS 17020. This corroborates 
the suggestion of the presence of dusty warm material in this galaxy
made on the basis of optical properties (Wisotzki \& Bade 1997, L97)
and the detection of an oxygen edge in the \asca spectrum (L97). 
This first good case for a {\em dusty} warm absorber in a NLSy1-type galaxy, 
together with the other good candidates in the Sy\,1.8 galaxy NGC 3786
(KoFi 1997c),
the Sy\,1.5 NGC 3227 (KoFi 1996,1997b), the Sy\,1 MCG 6-30-15
(Reynolds et al. 1997), and the quasar IRAS 13349 (Brandt et al. 1996)
suggests this component to be common in all types of AGN.     

Given the strong X-ray absorption edges of neutral dust-bound metals, 
dusty warm absorbers will play an important role
not only in probing components of the active nucleus, like the dusty torus 
that is invoked in unification schemes,  
but also are they a very useful diagnostic of the (otherwise hard to determine)
dust properties
and dust composition in other galaxies. 
Current and future X-ray missions with sensitivity at soft energies and high 
spectral resolution  
(like SAX, AXAF and Spektrum-X) will play an important
role in studying these issues.      
 
\begin{acknowledgements}
The \ros project is supported by the German Bundes\-mini\-ste\-rium
f\"ur Bildung, Wissenschaft, Forschung und Technologie (BMBF/DLR) and the Max-Planck-Society.
We thank Gary Ferland for providing {\em{Cloudy}}, and Hartmut Schulz, Jochen Greiner
and Dirk Grupe  
for a critical reading of the manuscript.
\end{acknowledgements}

\end{document}